\documentstyle[amstex,amssymb,12pt]{article}
\newcommand{\beq}{\begin{equation}}
\newcommand{\eeq}{\end{equation}}
\newcommand{\ba}{\begin{array}}
\newcommand{\ea}{\end{array}}
\newcommand{\bea}{\begin{eqnarray}}
\newcommand{\eea}{\end{eqnarray}}
\newcommand{\bean}{\begin{eqnarray*}}
\newcommand{\eean}{\end{eqnarray*}}

\newtheorem{theorem}{Theorem}[section]
\newtheorem{prop}[theorem]{Proposition}

\newtheorem{defi}[theorem]{Definition}
\newtheorem{remark}[theorem]{Remark}
\newenvironment{rem}{\begin{remark} \rm}{\end{remark}}
\newtheorem{proof}{Proof.}
\newenvironment{pf}{\begin{proof} \rm}{\hfill$\square$ \end{proof}}

\newcommand{\CW}{{\cal W}}
\newcommand{\CU}{{\cal U}}
\newcommand{\CV}{{\cal V}}

\newcommand{\CN}{{\cal N}}
\newcommand{\CM}{{\cal M}}

\newcommand{\CP}{{\cal P}}
\newcommand{\CG}{{\cal G}}
\makeatletter
\@addtoreset{equation}{section}

\makeatother
\newcommand{\HH}{{\Bbb H}}

\newcommand{\CC}{{\Bbb C}}

\newcommand{\ZZ}{{\Bbb Z}}

\newcommand{\NN}{{\Bbb N}}

\newcommand{\WW}{{\Bbb W}}

         \def\Ga{\Gamma}

\def\la{\lambda}        \def\La{\Lambda}

\def\sig{\sigma}

\newcommand{\cmp}[3]{Comm. Math. Phys. {\bf #1} (#2), #3}

\newcommand{\jmp}[3]{J. Math. Phys. {\bf #1} (#2), #3}
\newcommand{\rref}[1]{(\ref{#1})} 

\def\dsl#1{{\displaystyle #1}}

\def\plusminus{\pm}
\newcommand{\del}{{\partial}}

\def\mapleft#1{\smash{\mathop{\longleftarrow}
\limits^{#1}}}

\def\Fdb{{Fa\`a di Bruno}}

\def\parpo#1#2{\{#1,#2\}}
\def\dpt#1#2{\frac{\partial #1}{\partial t_{#2}}}

\def\H#1{H^{(#1)}}

\def\W#1{W^{(#1)}}

\def\h#1{h^{(#1)}}

\def\endpf{\par\hfill$\square$\par\medskip\par\noindent}

\def\res{\mbox res}

\def\ger{hierarch}
\def\var{manifold}
\def\bih{bihamiltonian}
\def\varb{\bih\ \var}
\def\ham{Hamiltonian}

\def\ger{hierarch}

\def\parpu{{\parpo{\cdot}{\cdot}}}

\def\varb{\bih\ \var}

\begin{document}
\begin{titlepage}
\begin{flushright}
Ref. SISSA 82/97/FM
\end{flushright}
\null\vspace{1.truecm}
\begin{center}
\baselineskip=36pt
{\LARGE Bihamiltonian Geometry,
Darboux Coverings, \\ \vspace{.3truecm} and 
Linearization  of the KP Hierarchy }
\end{center}
\vspace{0.8truecm}
\makeatletter
\begin{center}
{\large
Gregorio Falqui${}^1$,
Franco Magri${}^{2,4}$, and
Marco Pedroni${}^{3,4}$}\\
\vspace{0.2truecm}
${}^1$ SISSA, Via Beirut 2/4, I-34014 Trieste, Italy\\
E--mail: falqui@sissa.it\\\vspace{0.3truecm}
${}^2$ Dipartimento di Matematica, Universit\`a di Milano\\
Via C. Saldini 50, I-20133 Milano, Italy\\
E--mail: magri@vmimat.mat.unimi.it\\\vspace{0.3truecm}
${}^3$ Dipartimento di Matematica, Universit\`a di Genova\\
Via Dodecaneso 35, I-16146 Genova, Italy\\
E--mail: pedroni@dima.unige.it\\\vspace{0.3truecm}
${}^4$ Centre E. Borel, UMS 839 IHP,\\
(CNRS/UPMC)--Paris, France
\end{center}
\makeatother
\vspace{.5truecm}
\begin{abstract}\noindent
We use ideas of the geometry of bihamiltonian
manifolds, developed by Gel'fand and Zakharevich, to study the KP 
equations. In this approach they have the form of local conservation laws,
and can be traded for a system of ordinary differential equations of Riccati
type, which we call the Central System. We show that the latter can be
linearized by means of a Darboux covering, and we use this procedure
as an alternative technique to
construct rational solutions of the KP equations.
\end{abstract}
\vspace{2.2truecm}
Work supported by the Italian M.U.R.S.T. and by
the G.N.F.M. of the Italian C.N.R.
\end{titlepage}
\setcounter{footnote}{0}
\section{Introduction}
In this paper we study some aspects of the KP theory from 
the point of view of the bihamiltonian approach to integrable 
systems. Our purpose is twofold. At first we describe how the KP
theory can be defined by means of a suitable 
application of the method of Poisson pencils, 
in the light of the so--called Gel'fand--Zakharevich (hereinafter GZ)
theorem on the local geometry of a \bih\ 
manifold~\cite{GZ93}. Secondly we discuss, from this point of view,
how one can trade the KP hierarchy of partial differential 
equations for a system of ordinary differential equations, which
we term {\em Central System}. We show 
how this system can be linearized and solved
by means of a Darboux transformation.\par
Our approach is inductive.
We start from the KdV theory, which 
we tackle as the GZ theory of the Poisson pencil    
\begin{equation}\label{gzp}
P_\la=-\frac12 \del^3_x+2(u+\la)\del_x+u_x,
\end{equation}
defined on the manifold of $C^\infty$--functions on $S^1$. 
Following the GZ scheme we study the
Casimir function $H_\la$ of this pencil.  We show that it
can be written in the integral form
\begin{equation}
H(z)=2z \int_{S^1} h(x,z) dx
\end{equation} 
where the local density $h$ is a Laurent series 
\begin{equation}
\label{accagen}
h=z+\sum_{j\ge 1} h_j z^{-j}
\end{equation}
in  $z=\sqrt{\la}$.
This density is 
related to the point $u$ of the phase space by 
the Riccati equation~\cite{MGK68,Ku81} 
\begin{equation}\label{riccat}
u+z^2=h_x +h^2,
\end{equation}
which, as is well known~(see, e.g.,~\cite{Alber}) defines all the 
coefficients $h_j$, $j\ge 1$, as differential polynomials of $u$.\par
The next step is to study the conservation laws associated with the 
KdV hierarchy.
A trivial but far reaching consequence of the involutivity
of the KdV Hamiltonians
is that the KdV flows 
imply the
local conservation 
laws~\cite{Wil81,Ch78,CFMP4}
\begin{equation}\label{ourkp}
\dsl{\dpt{}{j}} h=\del_x\H{j}
\end{equation}
for the density $h$.\par
These equations introduce the principal characters of our picture:
the currents $\H{j}$.
We show that they can be computed in the following way. Among the 
(finite) linear combinations $\sum C_l\h{l}$
of the \Fdb\ iterates 
\begin{equation}
\h{j}=(\del_x+h)\h{j-1},\qquad \h{0}=1,
\end{equation}
of $h$, with coefficients $C_l$ independent of $z$, we
pick up, for every $j\ge 0$, the unique combination
having the asymptotic expansion
\begin{equation}
\H{j}=z^j+\sum_{k\ge 1} H^j_k z^{-k}.
\end{equation}
When we insert these currents into the conservation laws~\rref{ourkp}
we obtain a hierarchy of mutually commuting vector fields on a generic 
Laurent series of the type~\rref{accagen}.
They are (a possible form of) the celebrated
KP equations.
When $h$ is required to be a solution of the Riccati equations~\rref{riccat},
these equations collapse into the conservation laws 
associated with the Poisson pencil~\rref{gzp}. 
\par
As a further step we study the time evolution of the 
currents $\H{j}$.
We show that they  satisfy a closed system of
ordinary differential equations which has the form of a generalized
Riccati system:
\begin{equation}
\frac{\partial H^{(k)}}{\partial t_j}=H^{(j+k)}-H^{(j)}H^{(k)}+\sum_{l=
1}^kH^j_lH^{(k-l)}+
\sum_{l=1}^jH^k_lH^{(j-l)}.
\end{equation}
This we call the Central System (CS). 
It encompasses and extends the KP hierarchy.
In~\cite{CFMP5} we have discussed how the KP equations can be recovered
from CS by a projection on the orbit space of the first vector
field $\dsl{\dpt{}{1}}$. Different projections allow to obtain those
KP systems related to  fractional KdV hierarchies~\cite{BaDe91,dGHM92}.\par
Finally we turn to the problem of solving
the Central System. By means of the method of Darboux covering, 
discussed in~\cite{mpz}, we prove that the Miura--like map
\begin{equation}\label{lamborgh}
\H{j}=\left(\sum_{l=0}^{j} W^0_{j-l} \W{l}\right)/ \W{0}
\end{equation} 
connects the Central System with another system of Riccati 
equations, defined on the
space of sequences of Laurent series $\{\W{k}\}_{k\ge 0}$ 
of the form
\begin{equation}
\W{k}=z^k+\sum_{l\ge 1} W^k_l z^{-l},
\end{equation}
which reads
\begin{equation}
\dsl{\dpt{}{j}} \W{k}+ z^j \W{k}=\W{j+k}+\sum_{l=1}^j W_l^k \W{j-l}.
\end{equation}
This system 
(which we call the {\em Sato System}, see \cite{Ta89})
can be explicitly linearized using 
methods well known from the theory of
Riccati equations~\cite{Reid}.\par
As a result the Miura map~\rref{lamborgh},  which in the present 
picture is the analog of a dressing transformation,
allows to construct explicit families of solutions of the 
Central System, and hence of the KP \ger y.
\par
In our opinion, this paper clarifies some issues in the analysis
of  the different pictures of the KP hierarchy, 
their mutual relations, and the linearization of the KP flows.
More specifically we refer to the following three representations
for KP:
\begin{itemize}
\item[a)] The Lax representation
in the space of pseudodifferential 
operators;
\item[b)]
The Sato representation as  linear flows 
of a maximal torus in $GL_\infty$ on the Universal Sato Grassmannian $UGr$;
\item[c)]
The ``bihamiltonian representation'' as conservation laws \rref{ourkp} 
satisfied by the Hamiltonian density $h$.
\end{itemize}    
The analysis of 
the representations a) and b), and of their
equivalence, has been deeply expounded
in a number of nowadays classical papers and lecture
notes (see, e.g.,~\cite{DJKM,DikBook,Ma78,VM91,Mu94,SS,SW85,Ta89}),
while the picture of the KP equations as conservation laws, although
already introduced in~\cite{Ch78,Fl83,SS,Wil81}, 
has somehow been left aside from the main 
stream of the research work on the subject.
By fully developing the \bih\ approach, this paper 
aims to show that the KP theory
can also be approached on a traditional ground, in the spirit of geometrical 
methods of classical mechanics~\cite{Arno}.
\par
The paper is organized as follows. 
It starts with a brief {\em r\'esum\'e} of the 
\bih\  theory, devoted only to those aspects of the GZ theorem
relevant to the paper.
The next three sections provide the description of the path from KdV to KP 
and the Central System, briefly discussed above.
In Section~\ref{sec4}
we consider the action of Darboux transformations 
on the Central System.
In the final sections we take advantage of such a point of view
to linearize the KP theory and we address the problem of
writing explicit solutions, which are Hirota--like polynomial solutions.
\section{The method of Poisson pencils (to construct integrable Hamiltonian 
system)}\label{sec0}
In the simplest setting of this method one considers a Poisson manifold
$\CM$ and a vector field $X$. The vector field is used 
to deform the 
Poisson bracket $\parpu$ on $\CM$. We denote by 
\begin{equation}
\begin{array}{l}
\parpo{f}{g}^\prime= \parpo{X(f)}{g}+ 
\parpo{f}{X(g)}-X(\parpo{f}{g})\\
\parpo{f}{g}^{\prime\prime}= \parpo{X(f)}{g}^\prime+ \parpo{f}{X(g)}^\prime
-X(\parpo{f}{g}^\prime)
\end{array}
\end{equation}
the first two Lie derivatives of this bracket along $X$.
As the unique condition on $X$, we demand that the second derivative 
identically vanishes on $\CM$:
\begin{equation}
\parpo{f}{g}^{\prime\prime}\equiv 0.
\end{equation}
In this case, the pull--back
\begin{equation}
\parpo{f}{g}_\la:=
\parpo{f\circ\phi_{-\la}}{g\circ\phi_{-\la}}\circ\phi_{\la}
\end{equation}
of the given bracket with respect to the flow $\phi_\la:\CM\to\CM$ 
associated with $X$ depends linearly on $\la$,
\begin{equation}\label{ppen}
\parpo{f}{g}_\la=\parpo{f}{g}-\la \parpo{f}{g}^\prime,
\end{equation}
and, therefore, it defines a linear pencil of Poisson brackets.
Under these circumstances, we say 
that $\CM$ is an 
(exact) \varb\  and that $X$ is its Liouville vector field.
The names are chosen to suggest the analogy with the case
of exact symplectic manifolds.\par
The basic idea of the method is to use the {\em Casimir functions\/} 
of the pencil~\rref{ppen} 
to construct integrable Hamiltonian systems on $\CM$. 
We describe this technique
in the case of an odd--dimensional manifold endowed with a Poisson 
pencil of maximal rank. 
This entails that $\parpu_\la$ has a unique Casimir function $H_\la$,
depending on $\la$.
Let us set $\dim \CM=2 n+1$. 
Gel'fand and Zakharevich~\cite{GZ93} have shown that $H_\la$ is a degree $n$
polynomial in $\la$, 
\begin{equation}
H_\la=H_0\la^n+H_1\la^{n-1}+\cdots+H_n,
\end{equation}
which starts with the Casimir function $H_0$ of $\parpu^\prime$
and ends with the Casimir function $H_n$ of $\parpo{\cdot}{\cdot}$.
The coefficients $H_j$ 
verify the recursion relations
\begin{equation}
\parpo{\cdot}{H_{j+1}}^\prime=\parpo{\cdot}{H_{j}},
\end{equation}
and therefore are
in involution with respect to all the brackets of the pencil:
\begin{equation}\label{invo}
\parpo{H_j}{H_k}_\la=0.
\end{equation}
In the compact case, their level surfaces are $n$--dimensional tori 
defining a
Lagrangian foliation of $\CM$.\par
To convert this result in a statement on dynamical systems, we 
consider the pencil of vector fields
\begin{equation}\label{1.6}
X_\la(f):=\parpo{f}{H_\la}^\prime
\end{equation}
associated with $H_\la$ through the deformed bracket $\parpu^\prime$. 
We make two remarks. First 
we notice that $X_\la$  is a bihamiltonian vector field since we can write
\begin{equation}\label{canhier}
X_\la(f)=\parpo{f}{H_\la}^\prime=\parpo{f}{H_\la^\prime}_\la.
\end{equation}
The derivative
\begin{equation}
H^\prime_\la=X(H_\la)
\end{equation} 
is the {\em second Hamiltonian function}. 
Then we notice that $X_\la$ is 
a completely integrable system in the sense of Liouville since
\begin{equation}\label{terfor}
X_\la(H_j)=0.
\end{equation}
We call the polynomial family of vector fields,
\begin{equation}\label{1.8}
X_\la= X_0\la^n+X_1\la^{n-1}+\cdots+X_n,
\end{equation}
the {\em canonical hierarchy\/} defined on the exact \varb\ $\CM$.

\section{The KdV hierarchy}
\label{sec1}
In this section we define the KdV hierarchy as the canonical hierarchy
on a special exact \bih\ \var, and we use this point of view  
to pave the way to the KP theory.\par
In this example the manifold $\CM$ is the space  of 
scalar--valued $C^\infty$--functions on $S^1$, 
the Liouville vector field is
\begin{equation}
\dot u:=X(u)=1,
\end{equation}
and the Poisson pencil
is given in the form of a one--parameter family of skew--symmetric maps from
the cotangent to the tangent bundle~\cite{DikBook,CMP}: 
\begin{equation}\label{poipen}
\dot u=(P_\la)_uv=-\frac12 v_{xxx}+2(u+\la)v_x+u_xv.
\end{equation}
In this formula $u$ is a point of $\CM$, $v$ is a covector
attached at $u$, and the value of $v$ on a generic tangent 
vector $\dot u$ is given by
\begin{equation}
\langle v,\dot u\rangle=\int_{S^1}v(x)\dot u(x)\,dx,
\end{equation}
where $x$ is the coordinate on $S^1$.\par 
The first problem is to compute the Casimir 
function $H_\la$ and 
its derivative $H_\la^\prime$
along $X$.
In the present infinite--dimensional 
context
they can be conveniently  written as integrals 
\begin{eqnarray}
H &=& 2z\int_{S^1}h\,dx\\
H^\prime &=&\int_{S^1}h^\prime\,dx
\end{eqnarray}
of
local densities 
\begin{eqnarray}
h(z)=z+\sum_{j\geq 1}h_jz^{-j}\label{acca}\\
h^\prime(z)=1+\sum_{j\geq 1}h^\prime_jz^{-j},\label{accast}
\end{eqnarray}
which are Laurent series (rather than polynomials) in $z=\sqrt{\la}$.
\begin{prop}\label{p2.1}Let $h$ and $h^\prime$ be the unique
solutions of the Riccati
system 
\begin{eqnarray}
h_x+h^2=u+z^2\label{rich}\\
-\frac12 h^\prime_x+h h^\prime=z\label{richst}
\end{eqnarray} 
admitting the asymptotic expansions~\rref{acca} and~\rref{accast}.
Then their integrals $H_\la$ and $H_\la^\prime$ are respectively
the Casimir function of the Poisson pencil~\rref{poipen} 
and its associated
second Hamiltonian function.  
\end{prop}
{\bf Proof.} 
We use the identity
\begin{equation}
v\left(-\frac12v_{xxx}+2(u+z^2)v_x+u_xv\right)=\frac{d}{dx}\left(\frac14
v^2_{x}-\frac12vv_{xx}+(u+z^2)v^2\right)\label{3.8}
\end{equation}
to prove that the solution of the equation
\begin{equation}
\frac14 v^2_{x}-\frac12vv_{xx}+(u+\la)v^2=\la\label{3.9}
\end{equation}
belongs to the kernel of~\rref{poipen}.
Setting $\la=z^2$ we note that equation~\rref{3.9}
can be written in the form of a Riccati equation,
\begin{equation}
\left(\frac{z}{v}+\frac{v_x}{2v}\right)_x+
\left(\frac{z}{v}+\frac{v_x}{2v}\right)^2=u+z^2,\label{3.10}
\end{equation}
on
\begin{equation}
h(z):=\frac{z}{v}+\frac{v_x}{2v}.\label{3.11}
\end{equation}
By deriving this equation along any curve $u(t)$ in $\CM$,
we have $\dot u={\dot h}_x+
2h\dot h$. Therefore,
\begin{equation}
\langle v,\dot u\rangle=\int_{S^1}v(x)({\dot h}_x+
2h\dot h)\,dx = 2\int_{S^1}(-\frac12 v_x+hv)\dot h\,dx =
\frac{d}{dt}\left(2z\int_{S^1}h\,dx\right). 
\label{3.14}
\end{equation}
This formula proves that $v$ is the 
differential of the first Hamiltonian
$H_\la$, which, consequently, is the Casimir function 
we were looking for. 
To compute  the second \ham\ $H^\prime_\la$,
it suffices to notice that
\begin{equation}
H^\prime_\la=X(H_\la)=\langle v, 1
\rangle=\int_{S^1}v\,dx.
\end{equation}
This suggests to set 
\begin{equation}
h^\prime=v.
\end{equation}
Equations~\rref{rich} and~\rref{richst} follow from~\rref{3.10} 
and~\rref{3.11} respectively.
\endpf 
The next problem is to study the
canonical hierarchy associated with $H_\la$. 
It admits three different  representations,
according to the use of equations~\rref{1.6},~\rref{canhier}, or~\rref{terfor}
of Section~\ref{sec0}.
In the first representation,
based on formula~\rref{1.6}, the KdV equations are written as 
Hamiltonian equations
\begin{eqnarray}
\dsl{\dpt{u}{2j}}&=&-2 \del_x\, v_{2j+1}= 0\\
\dsl{\dpt{u}{2j+1}}&=&-2 \del_x\, v_{2j+2}
\end{eqnarray}
with respect to 
the derived bracket $\parpo{\cdot}{\cdot}^\prime$.  
In the second representation, based on~\rref{canhier},
they are
written as Hamiltonian equations 
with respect to the pencil $P_\la$. 
After some straightforward computations~\cite{Alber,CMP}, 
we get
\begin{eqnarray}
\dpt{u}{2j}&=& 0\label{2.5}\\
\dpt{u}{2j+1}&=&
\left(-\frac12\del^3_x+2(u+z^2)\del_x+u_x\right)\cdot(\la^j v(\la))_+,
\label{2.4}
\end{eqnarray}
where
\begin{equation}
(\la^j v(\la))_+=\sum_{i=0}^j v_{j-i}\la^i.
\end{equation}
In the third  representation, based on formula~\rref{terfor},
the attention is focused on the local Hamiltonian
density $h(z)$.  
It must obey local
conservation laws of the form 
\begin{equation}
\label{kpred}
\dpt{h}{j}=\del_x \H{j},
\end{equation}
where the $\H{j}$ are suitable ``current densities'',
since the Hamiltonian 
$H_\la$ is constant along the flows of the KdV \ger y.
The final problem is 
to compute these densities.
\begin{prop}\label{corrkdv}
The current densities 
$\H{j}$ of the KdV \ger y are given by the formulas
\begin{eqnarray}
\H{2j}&=&\la^j \label{evkdvcur}\\
\H{2j+1}&=&-\frac12{(\la^j v)_+}_x+h(\la^jv)_+
\label{kdvcur}.
\end{eqnarray}
\end{prop}
{\bf Proof.}
From equation~\rref{rich} we have
\begin{equation}
\dpt{u}{j}=(\del_x+2h)\left(\dpt{h}{j}\right),
\end{equation}
and from the second representation~\rref{2.4}
of the KdV equations we deduce
\begin{equation}
\dpt{u}{2j+1}=(\del_x+2h)(\frac12\del_x)(-\del_x+2h)(\la^j v)_+,
\end{equation}
by noticing that
\begin{equation}
-\frac12 v_{xxx}+2(u+\la)v_x+u_xv=(\del_x+2h)(\frac12\del_x)
(-\del_x+2h)\cdot v.
\end{equation}
Therefore
\begin{equation}
\begin{array}{rl}
(\del_x+2h)\left(\dsl{\dpt{h}{2j+1}}\right)&=
(\del_x+2h)(\frac12\del_x)\left(-{(\la^j v)_+}_x+2h(\la^jv)_+\right)\\
&=(\del_x+2h)\del_x\left(-\frac12{(\la^j v)_+}_x+h(\la^jv)_+\right),
\end{array}
\end{equation}
proving~\rref{kdvcur}.
The formula $\H{2j}=z^{2j}$ is obvious, since the even 
times are trivial.
\endpf
The definition~\rref{poipen} of the Poisson pencil,
the Riccati system~(\ref{rich}, \ref{richst}) for the Hamiltonians, and the 
definition~(\ref{kdvcur}, \ref{evkdvcur}) of the currents
are the basic formulas of the \ham\ theory of the KdV 
equations.
They introduce a new object, the currents $\H{j}$.
Their study is the leading theme of the paper.

\section{The KP hierarchy}\label{sec2}
The aim of this section is to give 
a new characterization of the current $\H{j}$
in terms of the \ham\ density $h(z)$. 
To this end we write \rref{kdvcur} in the equivalent 
forms
\begin{eqnarray}
\H{2j+1}&=z^{2j}(-\frac12 v_x+hv)+\frac12 (z^{2j} v)_{-\;,x}-
h(z^{2j}v)_-
\label{3.1}\\
\H{2j+1}&=\sum_{l=1}^j\left[-\frac12 v_{j-l\;,x}(z^{2l}\cdot 1)+v_{j-l}
(z^{2l}\cdot h)\right],\label{3.2}
\end{eqnarray}
where $(z^{2j} v)_-$ denotes the strict negative part of the expansion
of $z^{2j} v$ in powers of $z$. Each of these representations 
points out an important 
property of the currents $\H{j}$.
Equation \rref{3.1} allows 
to control the expansion of $\H{2j+1}$ in powers of $z$. Indeed, from 
it we obtain 
\begin{equation}
\H{j}=z^j+O(z^{-1})\label{zexp}
\end{equation}
by noticing that the second Riccati equation~\rref{richst} implies 
$z^{2j}(-\frac12 v_x+hv)=z^{2j+1}$. The interpretation of \rref{3.2} is more 
subtle: it provides a different type of expansion of the current $\H{2j+1}$
on a basis attached to the Hamiltonian density $h$. To display this 
expansion, we consider
the \Fdb \ iterates of $\h{0}=1$ at the point $h$, defined by
\begin{equation}
\h{j+1}=(\del_x+h)\cdot \h{j}.
\end{equation}
The linear space spanned by them (over $C^\infty$--functions) is 
denoted by $H_+$. Since 
\begin{equation}
\h{2}=h_x+h^2,
\end{equation}
we can write the Riccati equation~\rref{rich} in the form
\begin{equation}
z^2=\h{2}-u\h{0},
\end{equation}
showing that $z^2\in H_+$. 
Applying the operators $(\del_x+h)^j$ to both sides 
of this equation, one shows that 
$z^2(H_+)\subset H_+$. In particular, $z^{2j} \cdot 1\in H_+$ and $z^{2j}\cdot
h\in H_+$ for $j\ge 0$. Then equation \rref{3.2} means that the 
currents $\H{2j+1}$ belong to $H_+$. The same is trivially true for $\H{2j}$.
Therefore we conclude that all the currents $\H{j}$ are {\em \Fdb\ 
polynomials\/}
\begin{equation}\label{fdbexp}
\H{j}=\sum_{l=0}^j c_l^j \h{l}
\end{equation}
with coefficients $c_l^j$ independent of $z$. For any $j\ge0$
there is a unique choice of these coefficients leading to a ``degree''
$j$ \Fdb \ polynomial with the asymptotic expansion~\rref{zexp} 
in power of $z$:
the currents $\H{j}$ are (with alternating signs)
the principal minors of the infinite matrix
\begin{equation}
{\HH}=\left[
\begin{array}{ccccc} h^{(0)}& h^{(1)}& h^{(2)}  &\h{3}&\dots\\
    \res\frac{h^{(0)}}{z}& \res \frac{h^{(1)}}{z}
&\res \frac{h^{(2)}}{z} & \res \frac{h^{(3)}}{z}&\dots\\
0 &\res\frac{h^{(1)}}{z^2}& \res \frac{h^{(2)}}{z^2}&\res \frac{h^{(3)}}{z^2}& \dots\\
 ~ & 0 & \res\frac{h^{(2)}}{z^3}&\res\frac{h^{(3)}}{z^3}& \dots\\
 ~ & ~ & 0 & \res\frac{h^{(3)}}{z^4}&\dots\\
\dots & \dots & \dots & \dots&\dots\end{array}\right].
\label{hmat}
\end{equation}
Summarizing:
\begin{prop}\label{proex}
The current densities of the KdV theory 
are the principal minors of the matrix~\rref{hmat}, i.e., they are the
unique \Fdb\ polynomials~\rref{fdbexp} having the
asymptotic expansion~\rref{zexp} in powers of $z$.
\end{prop}
The advantage of this definition with respect to the one of Proposition
~\rref{corrkdv} is that 
it does no longer require that the Laurent series
\begin{equation}
\label{monh}
h(z)=z+\sum_{l\ge 1} h_l z^{-l}
\end{equation}
be a solution of the Riccati equation \rref{rich}. We have already 
encoded the \ham\ 
origin of the currents $\H{j}$ into the \Fdb\ expansion \rref{fdbexp}. 
We can thus forget the Poisson pencil~\rref{poipen} and the associated 
Riccati system, and retain simply the property stated in  
Proposition~\ref{proex}. Henceforth, we shall regard it as the definition
of the currents $\H{j}$ associated with {\em any} monic Laurent series 
\rref{monh}. This allows to extend equations \rref{kpred} to these 
series.
\begin{defi}\label{kpdef}
The KP equations are the equations
\begin{equation}\label{kpfl}
\dpt{}{j} h = \del_x \H{j}
\end{equation}
on an arbitrary monic Laurent series \rref{monh},
where the currents $\H{j}$ are the \Fdb\ polynomials considered
in Proposition~\ref{proex}.
\end{defi} 
After a suitable 
change of variables~\cite{CFMP4},
this definition reproduces the standard one, usually written 
in the language of pseudodifferential operators (see, 
e.g.,~\cite{DJKM,DikBook}). Let us briefly explain this relation. 
First of all, we consider the negative \Fdb\ iterates, obtained
by solving 
backwards the recursion relations
\begin{equation}
\h{j+1}=(\del_x+h)\h{j},\qquad\qquad j<0.
\end{equation}
The coefficients of the $\h{j}$ can be computed recursively, and one 
can easily show that $\h{j}=z^j+O(z^{j-1})$. Then we develop $z$ on 
the basis $\{\h{j}\}_{j\in\Bbb Z}$:
\begin{equation}\label{zfunh}
z=h-\sum_{j\geq 1}q_jh^{(-j)}.
\end{equation}
This gives an invertible relation between the coefficients $h_i$ of 
$h$ and the $q_i$.
For instance, the first relations are
\begin{equation}
\begin{array}{l}
q_1=h_1,\quad
q_2=h_2,\quad
q_3=h_3+h_1^2,\\
q_4=h_4+3h_1h_2-h_1h_{1x}.
\end{array}
\end{equation}
Finally, we introduce the pseudodifferential operator 
\begin{equation}\label{laxop}
Q=\del-\sum_{j\geq 1}q_j\del^{-j}.
\end{equation}
One can show~\cite{CFMP4} that the KP 
equations \rref{kpfl} on $h$ entail the 
Lax equations on $Q$,
\begin{equation}\label{laxfl}
\dpt{Q}{j}=[Q,(Q^j)_+],
\end{equation}
where $(Q^j)_+$ is the purely differential part of the $j$--th power 
of $Q$. 
\par The 
transformation~\rref{zfunh} can be usefully compared with 
the change of representation in the classical treatment 
of the equations of motion of the Euler top. 
If we use the space representation 
we simply write the conservation law of the angular momentum as
\begin{equation}
{\frac{d L}{dt}}=0.
\end{equation}
If we use the body representation, we write the Euler equation
\begin{equation}
{\frac{d L}{dt}}=[L,\Omega].
\end{equation}
The same happens in the KP theory.
When we pass from the \ham\ representation~\rref{kpfl}
to the Lax representation~\rref{laxfl} we are performing 
the analog of the passage from the space representation to 
the body representation of classical mechanics.\par
We will not discuss the KP equations any longer, 
but rather consider them as an intermediate step towards
the main topic of this paper, i.e., the analysis of the equations on the 
currents $\H{j}$ themselves. 
\section{The Central System}\label{secfive}
In this section we shall see the Riccati equations appear again
in a disguised form. They arise here from the study of the time 
evolution of the currents $\H{j}$. We interpret 
the KP equations~\rref{kpfl}
as the commutativity conditions of the
operators $\partial_x +h$ and $\dsl{\frac{\partial}{\partial t_j}}+H^{(j)}$:
\begin{equation}
\left[\partial_x +h,\dpt{}{j}+H^{(j)}\right]=0.
\end{equation}
Since $\H{j}\in H_+$ and $H_+$ is invariant with respect to the
operator $\partial_x +h$, we see that  $H_+$ is invariant 
also with respect to the operators $\dsl{\dpt{}{j}}+H^{(j)}$,
\begin{equation}\label{invcon}
\left(\frac{\partial}{\partial t_j}+H^{(j)}
\right)(H_+)\subset H_+,
\end{equation}
as shown by the following simple argument:
\begin{equation}
\left(\dpt{}{j}+\H{j}\right)\h{k}=
\left(\dpt{}{j}+\H{j}\right)\left(\del_x+h\right)^k1=
\left(\del_x+h\right)^k \H{j}\in H_+.
\end{equation}
Let us now remark that the sequence  $\{\H{j}\}_{j\ge 0}$ is, in its turn,
a basis in $H_+$. Then the previous invariance condition implies that there
exist coefficients $\gamma^{jk}_l$ (independent of $z$) such that 
\begin{equation}\label{levi}
\frac{\partial H^{(k)}}{\partial t_j}+H^{(j)}H^{(k)}=
\sum_{l=0}^{j+k} \gamma^{jk}_l(h)H^{(l)}.
\end{equation}
They can be easily identified by comparing the expansion of both
sides of this equation in powers of $z$. 
\begin{prop} Along the trajectories of the KP \ger y, the 
current densities $\H{k}$ obey the equations
\begin{equation}\label{2.8}
\frac{\partial H^{(k)}}{\partial t_j}+H^{(j)}H^{(k)}=H^{(j+k)}+\sum_{l=
1}^kH^j_lH^{(k-l)}+
\sum_{l=1}^jH^k_lH^{(j-l)},
\end{equation}
which we call the {\em Central System (CS)} 
associated with the KP theory. 
\end{prop}
In these equations the \ham\ density $h(z)$ plays no special role. 
Hence we can forget the \Fdb \ rule to construct the polynomials
$\H{j}$, and we can look at them simply as a collection
$\{\H{j}\}_{j\in\NN}$ of Laurent series,
\begin{equation}
\H{j}=z^j+\sum_{l\ge 1} H^j_l z^{-l},
\end{equation}
with independent coefficients $H^j_l$.
From this point of view, the Central System,
written in the componentwise form
\begin{equation}
\dpt{H_m^k}{j}+H_{m+k}^j+H_{j+m}^k+\sum_{l=1}^{m-1}H_l^j H_{m-l}^k
=H_m^{j+k}+\sum_{l=1}^{j-1}H_l^k H^{j-l}_m+\sum_{l=1}^{k-1} H_l^j H^{k-
l}_m,
\end{equation}
is manifestly a system of {\em ordinary\/} differential equations 
of Riccati type for the new variables $ H^j_l$. \par 
Through our two--steps process from the KdV equations
to the Central System, we eventually passed from a hierarchy of
partial differential equations to a dynamical system. This result,
which is originally due to Sato, admits an interesting geometrical 
interpretation~\cite{CFMP5}, which we briefly recall. The idea is that
the KP and KdV equations are ``reduction'' of the Central System,
and the problem is to understand this reduction process.
We have to recall that the vector 
fields of the Central System pairwise commute. Then we are allowed
to perform two kind of ``reductions''. The first is a restriction to
the submanifold of singular points of any vector field of the system.
The second is a projection onto the orbit space of any vector field
of the system along its trajectories. The two processes commute. 
In~\cite{CFMP5} we have shown that KP can be obtained from CS as the 
projection along the trajectories of the first vector field of CS.
It is this process which converts the original family of
ordinary differential equations into a \ger y of partial differential
equations. The projection is defined 
by the \Fdb \ condition  
\begin{equation}
\H{k}=\sum_{l=0}^k c_l^k \h{l}.
\end{equation}
On the other hand, as it is well known, KdV is a restriction
of KP on the manifold of singular points of the second vector field of the
\ger y. The restriction is defined by the Riccati equation
\begin{equation}
h_x+h^2=u+z^2.
\end{equation}
Therefore the passage from CS to KdV is a combined process,
involving both a projection and a restriction. This gives the geometrical
meaning of the \Fdb\ condition and of the Riccati equation. 
Of course this is only the simplest example of such a kind of procedure.
Other examples are the so--called fractional KdV \ger ies~\cite{dGHM92},
see~\cite{CFMP5}.\par\noindent
\begin{rem}
We end this section on the Central System with some cursory remarks on its
relation with the theory of the $\tau$--function and of the 
Baker--Akhiezer function $\psi$~(see,
e.g.,~\cite{DJKM,DikBook,Hir,VM91,SS,SW85}).\par 
The link with the Baker--Akhiezer function $\psi$ and the alternative forms of
the KP equations suggested by Sato and Sato in the seminal paper~\cite{SS}
rests on the following  argument.
A glance at the Central System shows the symmetry conditions
\begin{equation}\label{symmcond}
\dpt{\H{j}}{k}=\dpt{\H{k}}{j}.
\end{equation}
Therefore, there exists a function $\psi$ such that
\begin{equation}
\dpt{\psi}{j}=\H{j}\psi.
\end{equation}
With a suitable normalization this is a Baker--Akhiezer function.
Moreover, by the same conditions, the differential operators
\begin{equation}
D_j:=\dpt{}{j}+\H{j}
\end{equation}
commute.
By acting recursively with these operators on the 
lowest order current $\H{0}=1$ one obtains vectors 
$D_{j_1}\cdots D_{j_k}(\H{0})$ belonging to $H_+$ 
thanks to the invariance condition~\rref{invcon}.
One can see that they satisfy remarkable constraints.
For instance, we have, for $a,b\in\Bbb C$,
\bean
(a D_2+b D_1^2)(1)&=& a\H{2}+b(\H{2}+2H_1^1\H{0})\\
                  &=& (a+b)\H{2}+2b H_1^1\H{0},
\eean
so that
setting $a=-b$ we can force the above linear combination to be a 
multiple of $\H{0}$.
Analogously
\bean
(a D_3+b D_1 D_2+c D_1^3)(1)&=& (a+b+c)\H{3}+(b+3c)H_1^1\H{1}\\
                            &&   +[(b-3c)H_2^1+(b+3c)H_1^2]\H{0}
\eean
is independent of $z$ for $a=2c$ and $b=-3c$. 
In general it can be proved that the
vectors $D_{j_1}\cdots D_{j_k}(1)$ verify the constraints
\begin{equation}
\label{vacuum}
p_k(-\frac1{l}D_l)(1) = H_{k-1}^1,
\end{equation}
where the Schur polynomials $p_k(t_1,t_2,\dots)$ are as usual defined  
via the relation
\begin{equation}\label{schurpol}
\exp(\sum_{i\ge 1} t_i z^i)=\sum_{k\ge 0} p_k(t_l) z^k.
\end{equation}
Then the simple identity
\begin{equation}
\psi\cdot p_k(-\frac1{l}D_l)(1)=
p_k(-\frac1{l}\frac{\del}{\del t_l})(\psi)
\end{equation}
allows us to recover the Sato constraints
\begin{equation}
p_k(-\frac1{l}\frac{\del}{\del t_l})(\psi) = H_{k-1}^1\cdot \psi
\end{equation}
presented in \cite{SS}.\par
The link with the $\tau$--function is
provided by the second \ham\ function $H^\prime_\la$ discussed in Section
\ref{sec1}. It is preserved along the flows of the KP hierarchy too, and
therefore its \ham\ density $h^\prime$ verifies local conservation  
laws of the form
\begin{equation}
\frac{\del h^\prime}{\del t_j}={\del }_x H^\prime_{(j)},\label{cld}
\end{equation}
which we call {\em dual KP equations}. Setting 
\begin{equation}
H_{(l)}=z^{l-1}-\sum_{k\ge 1}H^k_l z^{-(k+1)},
\end{equation}
one finds~\cite{CFMP4} that the dual currents $H^\prime_{(j)}$ are given by 
\begin{equation}
H^\prime_{(j)}=\sum_{l=1}^j H_{(l)}\H{j-l}.
\end{equation}
This formula allows us to compute the coefficients $H^\prime_{jk}$
of the expansion of $H^\prime_{(j)}$ in powers of $z$,
\begin{equation}
H^\prime_{(j)}=jz^{j-1}-\sum_{l\ge1}H^\prime_{jl}z^{-(l+1)},\label{uno.20}
\end{equation}
as quadratic polynomials of the coefficients $H^i_k$ of
the primal currents $\H{i}$. By using this representation one can show
the symmetry property
\begin{equation}
H^\prime_{jk}=H^\prime_{kj}.
\end{equation}
Furthermore, as functions of the times $(t_1,t_2,\dots)$, they verify the
differential conditions 
\begin{equation}
\dpt{H^\prime_{jk}}{l}=\dpt{H^\prime_{lk}}{j}.
\end{equation}
Therefore, there exists a function $\tau(t_1,t_2,\dots)$ independent of $z$
such that 
\begin{equation}
H^\prime_{jk}=\frac{\del^2}{\del t_j\del t_k}\log\tau.
\end{equation}
This function is the Hirota $\tau$--function associated with the Central
System. As it is well known, by introducing this function it is possible to
set the KP \ger y in the form of Hirota bilinear equations. However, it is
also possible to set directly the equations of the Central System in the form
of a linear system by means of a suitable transformation, to be discussed in
the next sections.
\end{rem}

\section{Darboux Coverings and the Central System}
\label{sec4}
To linearize the Central System we shall use the 
method of Darboux coverings~\cite{mpz}.
The hallmark of such a rather ``unconventional'' 
approach  to the classical subject of Darboux maps and symmetries is quite
simple. In a first instance, one  replaces the search for a
transformation between two vector fields $X$ and $Z$ 
defined on the manifolds $\CM$ and $\CP$ respectively,
by that of a
a third vector field $Y$ (defined in general on a
bigger manifold $\CN$) separately related to $X$ and $Z$
by  two maps
$\pi:\CN\rightarrow \CM$ and $\sigma: \CN \rightarrow \CP$: 
\begin{equation}
X=\pi_*(Y),\qquad Z=\sigma_*(Y).
\end{equation}
By definition, integral curves of $Y$ are mapped to integral curves of 
$X$ by $\pi$, and to integral curves of $Y$ by $\sigma$.
We say for short that
Y {\em intertwines} $X$ with $Z$.
Moreover, we say that $Y$ is a {\em Darboux covering} of $X$
if $\CN$ is a fiber bundle over
$\CM$ and $\pi$ is the canonical projection. 
In this case,
any section $\rho$ of $\pi$, invariant under $Y$,
allows us to define a (Miura) map $\mu:\CM\to\CP$ 
relating directly the vector 
fields $X$ and $Z$. In pictures:
\[
\begin{array}{ccccccccccc}
&&Y&&&\qquad&&&Y&&\\
&\sig_*\swarrow&&\searrow{\pi_*}&&\qquad&&\sig_*\swarrow&&\nwarrow{\rho_*}&\\
Z&&&&X&\qquad&Z&&\mapleft{\dsl{\mu_*}}&&X
\end{array}\]
In the present instance, $\CM$ is the space of sequences of Laurent series
$\{\W{k}\}_{k\ge 0}$ of the form
\begin{equation}
\W{k}=z^k+\sum_{l\ge 1}W^k_l z^{-l}.
\end{equation} 
This space is a natural parameter space
for the big cell in the Sato Grassmannian~\cite{SS,SW85}.
The manifold $\CP$ is a second copy of $\CM$, formed by sequences
$\{\H{k}\}_{k\ge 0}$ of the form
\begin{equation}
\H{k}=z^k+\sum_{l\ge 1}H^k_l z^{-l}.
\end{equation}
Since the sequences $\W{k}$ and $\H{k}$ will play different 
roles in the sequel, it is convenient to regard the spaces $\CM$ and
$\CP$ as distinct. Finally,
the manifold $\CN$ is the Cartesian
product $\CM\times \CG$ of $\CM$ by the  
group $\CG$ of invertible Laurent series of the form
\begin{equation}
w=1+\sum_{l\ge 1} w_l z^{-l}.
\end{equation}
The vector field $Z$ on $\CP$ is
any vector field of the Central System~\rref{2.8}.
We recall that it is completely characterized by the property
\begin{equation} 
\label{sceqinva}
\left(\dpt{}{j} +\H{j}\right)H_+\subset H_+,
\end{equation}
where $H_+$ is the linear span of the Laurent series $\{\H{k}\}_{k\ge 0}$.
The vector field $X$ on $\CM$ is analogously characterized by the 
property
\begin{equation}\label{msceqinv} 
\left(\dpt{}{j}+ z^j\right)W_+\subset W_+,
\end{equation}
where $W_+$ is the linear span of the Laurent series $\{\W{k}\}_{k\ge 0}$.
By comparing the expansions of both sides of equation~\rref{msceqinv}
in powers of $z$, it is easily seen that the equations defining $X$ are
\begin{equation}
\label{msceq}
\dpt{}{j} \W{k}+ z^j \W{k}=\W{j+k}+\sum_{l=1}^j W_l^k \W{j-l}.
\end{equation}
It can be shown~\cite{Ta89} that these are precisely the linear
flows on the Sato Grassmannian.
Finally, to define the vector field $Y$ on $\CN$ we impose the further
invariance condition
\begin{equation}\label{wcinv} 
\left(\dpt{}{j}+ z^j\right)(w)\in W_+,
\end{equation}
which is tantamount to defining
\begin{equation}\label{weq}
\dpt{}{j}w +z^j w=\sum_{l=0}^j w_l \W{j-l}.
\end{equation}
We summarize this discussion in the following
\begin{defi}
The Central System (CS) is the family of vector fields
\[
\frac{\partial H^{(k)}}{\partial t_j}+H^{(j)}H^{(k)}=H^{(j+k)}+\sum_{l=
1}^kH^j_lH^{(k-l)}+
\sum_{l=1}^jH^k_lH^{(j-l)}
\]
on $\CP$ uniquely characterized by the invariance condition
\rref{sceqinva}.\par
The Sato System (S) is the family of vector fields
\[
\dsl{\dpt{}{j}} \W{k}+ z^j \W{k}=\W{j+k}+\sum_{l=1}^j W_l^k \W{j-l}
\]
on $\CM$ uniquely characterized by the invariance condition
\rref{msceqinv}.\par
The Darboux Sato System  (DS) is the family of vector fields
\[\begin{array}{rcl}
\dsl{\dpt{}{j}} \W{k}+ z^j \W{k}&=&\W{j+k}+\sum_{l=1}^j W_l^k \W{j-l}\\
\dsl{\dpt{}{j}} w + z^j w&=&\sum_{l=0}^j w_l \W{j-l}
\end{array}
\]
on $\CN$ uniquely characterized by the invariance conditions 
\rref{msceqinv} and \rref{wcinv}.
\end{defi}
To complete the geometrical scheme of  Darboux covering we have yet to
define the maps $\sigma$ and $\pi$. The map $\pi$ is of course the
canonical projection 
\begin{equation}
\pi(w,\{\W{k}\})=\{\W{k}\}.
\end{equation}
The map $\sigma$ is defined by imposing the intertwining 
condition 
\begin{equation}
w\cdot (H_+)=W_+
\end{equation}
on the linear spans $H_+$ and $W_+$. 
It means that multiplying any element $\H{j}$ by $w$ we get an element
of $W_+$. This happens if and only if 
\begin{equation}\label{4.7}
w \H{j}=\sum_{l=0}^j w_{j-l} \W{l}\quad\quad\quad \forall j\ge 0.
\end{equation}
\begin{defi} We say that 
the sequence $\{\H{k}\}_{k\ge 0}$ is related to the sequence
$\{\W{k}\}_{k\ge 0}$ by the
{\em Darboux transformation} generated by $w$,
and we write $H_+= D_w(W_+)$, if $w\cdot H_+= W_+$.
\end{defi}
We are now in a position to prove the main property of the
DS equations.
\begin{prop}
The DS system is a Darboux covering of the Sato System, 
intertwining it with CS.
\end{prop}
{\bf Proof.} The only thing to show is that 
$\sigma_*(\mbox{DS})=\mbox{CS}$. 
Notice that the definitions of $\sig$ and DS entail
\begin{equation}
\label{eqwh}
\dpt{w}{j}+z^j w =w \H{j}.
\end{equation}
This means that the operators 
$\left(\dsl{\dpt{}{j}} + z^j\right)$ and $\left(\dsl{\dpt{}{j}} + \H{j}\right)$
are intertwined by the multiplication by $w$, i.e.,
\begin{equation}
w\cdot \left(\dpt{}{j} + \H{j}\right)
=\left(\dpt{}{j} + z^j\right)\cdot w.
\end{equation}
Therefore
\begin{equation}
\begin{array}{rl}
w\cdot \left(\dsl{\dpt{}{j}} + \H{j}\right)(H_+) = & 
\left(\dsl{\dpt{}{j}} + z^j\right)(w(H_+))\\
= & \left(\dsl{\dpt{}{j}} + z^j\right) W_+\subset W_+,
\end{array}
\end{equation}
and consequently
$\left(\dsl{\dpt{}{j}} + \H{j}\right)(H_+)\subset H_+$, so that CS follows.
\endpf
We now exploit this result to define a {\em Miura map} relating directly 
S to CS. We consider in $\CP$ the submanifold
\begin{equation}
\H{0}=1,
\end{equation}
which is clearly invariant under CS, and we construct its inverse
image in $\CN$,
\begin{equation}\label{ww0}
w=\W{0},
\end{equation}
with respect to the Darboux map $\sig:\CN\to \CP$. The corresponding 
section $\rho:\CM\to\CP$ is given by $\rho(\{\W{k}\})=
(\W{0},\{\W{k}\})$.
\begin{prop}
The submanifold defined by equation~\rref{ww0} in $\CN$
is a section of $\pi:\CN\to \CM$ which is invariant under DS.
\end{prop}
{\bf Proof.} 
We have to compare the DS equations for the pair $(\W{0},w)$. They 
are: 
\begin{equation}\label{dmcs00}
\left\{
\begin{array}{l}
\dsl{\dpt{}{j}} w=-z^j w+\sum_{l=0}^jw_l \W{j-l}\\
\dsl{\dpt{}{j}} \W{0}=- z^j \W{0}+\sum_{l=0}^j W_l^0\W{j-l},
\end{array}\right.
\end{equation}
since $W^0_0=1$.
Hence
\begin{equation}
\dpt{}{j} (w-\W{0})= 
-z^j(w-\W{0})+\sum_{l=0}^j (w_l-W_l^0)\W{j-l},
\end{equation}
proving the statement.
\endpf
Motivated by this result, we give the following
\begin{defi}
The nonlinear map
\begin{equation}
\mu=\sigma\circ\rho:\CM\to\CP
\end{equation}
given by
\begin{equation}\label{hdarw}
\H{j}=\left(\sum_{l=0}^{j} W^0_{j-l} \W{l}\right)/ \W{0}.
\end{equation}
is the {\em Miura map} relating S to CS.
\end{defi}
It enjoys the property of mapping any solution of the Sato System 
into a solution of CS satisfying the constraint $\H{0}=1$.

\section{Linearization of the Sato System  
and families of solutions}\label{sec5}
The final step is to show that the Sato System
can be explicitly linearized. To this end
it is useful to write it in matrix
form. Notice that in components it reads
\begin{equation}\label{compmsc}
\dpt{W_m^j}{k} + W_{k+m}^j- W_m^{j+k}=\sum_{l=1}^k W_l^j W_m^{k-l}.
\end{equation}
If we consider the infinite shift matrix
\begin{equation}\label{lambda}
\La=\left[\begin{array}{ccccc}
0 & 1 & 0 & \cdots  & \\
0 & 0  &  1 & 0 &\cdots \\
\vdots & & \ddots & \ddots &\\
\vdots & & & \ddots & \ddots \\
\vdots & & &  & \ddots
\end{array}\right]
\end{equation}
and the convolution matrix of level $k$
\begin{equation}
\Gamma_k=\left[\begin{array}{cccccc}
0 &  \cdots & & 1 & 0 & \cdots\\
\vdots & & 1 & 0 & \cdots & \cdots\\
& \cdot & \cdot & & & \\
1 & 0 & & & & \\
\vdots & & & & &
\end{array}\right]
\end{equation}
we can write~\rref{compmsc} in the matrix form:
\begin{equation}\label{matmsc}
\dpt{}{k}\CW + \CW\cdot{}^T\La^k-\La^k\cdot\CW=\CW\Ga_k\CW.
\end{equation}
This equation belongs to a well-known class of linearizable 
matrix Riccati equations~\cite{Reid}.
\begin{prop}\label{prop5.1}
The infinite matrix $\CW$ is a solution of the matrix Riccati 
equation~\rref{matmsc} if and only if it has the form 
$\CW=\CV\cdot \CU^{-1}$, with $\CU$ and $\CV$ satisfying the
constant coefficients linear system
\begin{equation}\label{lsys}
\left\{
\begin{array}{cl}
\dsl{\dpt{}{k}} \CU=&{}^T\La^k \CU-\Ga_k\CV\\
\dsl{\dpt{}{k}} \CV=&\La^k\CV
\end{array}
\right.
\end{equation} 
\end{prop}
{\bf Proof.}  
We only have to check that {\em every} solution of 
equation~\rref{matmsc} can be obtained from \rref{lsys}. Let $\CW$ be 
such a solution, and let
\begin{equation}
\left\{
\begin{array}{cl}
\CV=& \exp(\sum_{k\ge 1}t_k\La^k)\\
\CU=& \CW^{-1}\CV
\end{array}
\right.
\end{equation} 
Then $(\CU,\CV)$ is a solution of \rref{lsys}.
\endpf
Since $\CU$ and $\CV$ are infinite matrices, in general one
cannot explicitly solve the linear system \rref{lsys}, and this
procedure would imply the discussion of suitable notions
of convergence for formal series in infinite variables.
Nevertheless, a lot of solutions can be constructed as 
follows\footnote{This is part of a joint work with J.P. Zubelli, 
which will appear in a forthcoming paper.}.
Let us notice that the constraints $W_i^j=0$ $\forall\ i>n$,
$j>m$, is compatible with equations~\rref{matmsc}. 
In other words, the space ${\Bbb W}_{m,n}$ of matrices
$\CW$ which have zero entries outside the first $m$ rows and the
first $n$ columns is invariant for the Sato System. If we denote with
${\CM}_{m,n}$ the $m\times n$ matrix obtained by the infinite
matrix $\CM$ by taking its $m\times n$ upper corner, then for
the matrices $\CW\in{\Bbb W}_{m,n}$ equations \rref{matmsc}
can be written as
\begin{equation}\label{matmscred}
\dpt{}{k}{\CW}_{m,n} + {\CW}_{m,n}\cdot\left({}^T\La_{n,n}\right)^k-
\left(\La_{m,m}\right)^k\cdot{\CW}_{m,n}={\CW}_{m,n}(\Ga_k)_{n,m}{\CW}_{m,n}.
\end{equation}  
These are matrix Riccati equations for {\em finite\/} matrices.
They can be linearized as in Proposition \ref{prop5.1}, and
explicitly solved. Their solutions depend only on
$\{t_k\}_{k=1,\ldots,m+n-1}$, and should be compared with the
Hirota polynomial solutions of the KP hierarchy.

\section{An explicit example}
To make more concrete the discussion about the finite rank solutions,
we present some explicit computations for the case of 
$\WW_{3,2}$. To avoid clumsy notations, let us 
redefine the matrix coefficients
appearing in~\rref{matmscred}
as follows:
\begin{equation}
{\CW}_{3,2}=W,\quad
\La_{3,3}=A,\quad
{}^T\La_{2,2}=B,\quad
(\Ga_k)_{2,3}=C_k,
\end{equation}
for $k=1,\dots,4$. Hence, the only non--vanishing coefficients are:
\begin{equation}
\begin{array}{c}
B=\left[\begin{array}{cc}0 & 0  \\ 1& 0 \end{array}\right],\quad
A=\left[\begin{array}{ccc}  0&1&0  \\ 0&0 &1   \\ 0  & 0 &0\end{array}\right],
\quad A^2,\quad C_1=
\left[\begin{array}{ccc}  1  & 0 & 0 \\  0 & 0 & 0
\end{array}\right],\\
C_2=\left[\begin{array}{ccc}  0  & 1 & 0 \\  1 & 0 & 0
\end{array}\right],\quad
C_3=\left[\begin{array}{ccc}  0  & 0 & 1 \\  0 & 1 & 0
\end{array}\right],\quad
C_4=\left[\begin{array}{ccc}  0  & 0 & 0 \\  0 & 0 & 1  \end{array}\right].
\end{array}
\end{equation}
This shows that the Sato System on  $\WW_{3,2}$ 
can be seen as a system of four
Riccati--type ordinary differential equations in $\CC^6$.
According to the recipe of Proposition~\ref{prop5.1},
we set 
$W=V U^{-1},$
where $V$ is a $3\times 2$ matrix, and $U$ is a 
non--singular matrix of rank $2$.
We study the Cauchy problems
\begin{equation}\label{exsys}
\left\{\begin{array}{l}
\dsl{\dpt{}{k}} V= A^k V\\
V(0)=W(0),
\end{array}\right.
\qquad\qquad
\left\{\begin{array}{l}
\dsl{\dpt{}{k}} U =B^k U -C_k V\\
U(0)={\mbox I}
\end{array}\right.
\end{equation}
We first consider the equation for $V$. Since $A$ is nilpotent,
the solution is the {\em polynomial}
\begin{equation}\label{solV}
V(t)=\exp(\sum_{l=1}^2 t_l A^l)  W(0).
\end{equation}
Now we recall the definition of
the Schur polynomials $p_l(t_1,t_2,\ldots)$ given in Equation ~\rref{schurpol}
and of their ``adjoint'' ones,
\begin{equation}
\exp(-\sum_{i=1}^\infty t_i z^i)=\sum_{l=0}^\infty \widetilde{p_l}(t_1,
t_2,\ldots) z^l.
\end{equation} 
Thus we can rewrite~\rref{solV} as
\begin{equation}
V(t)=\sum_{l=0}^2 p_l(t) A^l W(0)=
 \left[\begin{array}{ccc} 1 & p_1 & p_2  \\ 0 &1 & p_1 
\\ 0 & 0 & 1\end{array}\right]\cdot W(0),
\end{equation}
As far as the second set of equations in \rref{exsys} is
concerned, we put $U=\exp(t_1 B) U_0(t)$, so that these
equations can be rewritten as
\begin{equation}
\dsl{\dpt{}{k}} U_0=-(\sum_{l=0}^{1}\widetilde{p_l} B^l)\cdot
C_k(\sum_{j=0}^2 p_j A^j)\cdot W(0),
\end{equation}
with $U_0(0)={\mbox I}$. 
The Cauchy problems with initial data $U_0(0)={\mbox I}$ 
can thus be easily solved; we get
\begin{equation}
U_0(t)={\mbox I}-\left[\begin{array}{ccc} 
t_1 & t_2+\frac12 t_1^2 & t_3+t_1 t_2+\frac16 t_1^3\\ 
t_2-\frac12 t_1^2 & t_3-\frac13 t_1^3 & t_4+\frac12 t_2^2
-\frac12 t_1^2 t_2-\frac18 t_1^4
\end{array}\right] W(0).
\end{equation}
In particular, for
\[
W(0)=\left[\begin{array}{cc}0 & 0\\ 0 & 0 \\0 & 1
\end{array}\right]. 
\]
we obtain
\[
W(t)=V(t)U(t)^{-1}=\tau^{-1}\left[\begin{array}{cc}
-t_1t_2-\frac12 t_1^3 & t_2+\frac12 t_1^2\\
-t_1^2 & t_1\\
-t_1 & 1\end{array}\right],
\]
where $\tau=\det U(t)=1-t_4-\frac12 t_2^2+\frac12 t_1^2t_2+\frac18 
t_1^4$. Therefore the corresponding solution of S is
\begin{equation}
\begin{array}{l}
\W{0}=1+\tau^{-1}\left[-(t_1t_2+\frac12 t_1^3)z^{-1}+(t_2+\frac12 
t_1^2)z^{-2}\right]\\
\W{1}=z+\tau^{-1}\left[-t_1^2 z^{-1}+t_1 z^{-2}\right]\\
\W{2}=z^2+\tau^{-1}\left[-t_1 z^{-1}+ z^{-2}\right]\\
\W{k}=z^k\qquad\qquad\mbox{for }\ k\geq 3.
\end{array}
\end{equation}
Using the Miura map \rref{hdarw} we obtain a solution $\{\H{j}\}$ of
the Central System such that $\H{k}=z^k$ for $k=0$ and $k\geq 5$.
For example,  $\H{1}$ is given by
\begin{equation}
\W{0}\H{1}=\W{1}+W^0_1\W{0}.
\end{equation}
This means that the coefficients $H^1_k$ can be computed using the 
recursion relations 
\[
\begin{array}{l}
H^1_1=W^1_1+(W^0_1)^2-W^0_2\\
H^1_2=W^1_2+2 W^0_1 W^0_2-W^0_1 W^1_1-(W^0_1)^3\\
H^1_{j+2}=-(H^1_{j+1} W^0_1+H^1_j W^0_2)\qquad\mbox{for } j\ge 1.
\end{array}
\] 
In a more compact form,
\[
\begin{array}{l}
\H{1}=(\W{1}+W^0_1\W{0})/\W{0}\\
\phantom{\H{1}}=
-\tau^{-1}(t_1t_2+\frac12 t_1^3)+\dsl{\frac
{z+\tau^{-1}\left[-t_1^2 z^{-1}+t_1 z^{-2}\right]}
{1+\tau^{-1}\left[-(t_1t_2+\frac12 t_1^3)z^{-1}+(t_2+\frac12 
t_1^2)z^{-2}\right]}}\quad .
\end{array}
\]
As we have seen in Section \ref{secfive}, $h=\H{1}$ is a solution of 
the KP equations after putting $t_1=x$.

\section{Summary}
In this paper we have tried to give an overview of the \bih\ approach
to the KP theory, starting from the primitive idea of Poisson pencil 
to arrive to the polynomial solutions of these equations. The approach
consists of two parts, dealing with the equations and with their solutions
respectively.
In the first part we have traced the way from KdV to CS (through a double
process of extension), and backwards from CS to KdV (through a projection and
a restriction).
In the second part we have shown how to use the technique of Darboux coverings
to linearize the equations and, therefore, to construct explicit solutions.
We hope that, by providing an alternative view of the theory, the present
paper may clarify the logical structure of the \ham\ approach to
the KP theory.
\subsection*{Acknowledgments}
We thank P. Casati and J.P. Zubelli for illuminating discussions and 
comments. Two of us (F.M.\ and M.P.) would like to thank
the staff of the Centre Emile Borel and  the organizers  
of the semester {\em Integrable Systems\/}, 
Professors O. Babelon, P.\ van Moerbeke, and J.B. Zuber, for 
providing a warm environment where part of this 
work has been done. Thanks are also due to the anonymous referee
for useful remarks.

\end{document}